\documentclass[twocolumn]{aastex62}

\usepackage{natbib}
\usepackage{color}

\graphicspath{{./}}

\received{March 25, 2019}
\revised{May 11, 2019}
\accepted{May 18, 2019}
\submitjournal{ApJ}

\shorttitle{Discovery of an au-scale excess in millimeter emission of TW~Hya}
\shortauthors{Tsukagoshi et al.}

\begin{document}

\title{Discovery of an au-scale excess in millimeter emission from the protoplanetary disk around TW~Hya}

\correspondingauthor{Takashi Tsukagoshi}
\email{takashi.tsukagoshi@nao.ac.jp}

\author[0000-0002-6034-2892]{Takashi Tsukagoshi}
\affil{Division of Radio Astronomy, National Astronomical Observatory of Japan, Osawa 2-21-1, Mitaka, Tokyo 181-8588, Japan}

\author{Takayuki Muto}
\affil{Division of Liberal Arts, Kogakuin University, 1-24-2 Nishi-Shinjuku, Shinjuku-ku, Tokyo, 163-8677, Japan}

\author{Hideko Nomura}
\affil{Department of Earth and Planetary Sciences, Tokyo Institute of Technology, 2-12-1 Ookayama, Meguro, Tokyo, 152-8551, Japan}

\author{Ryohei Kawabe}
\affil{Division of Radio Astronomy, National Astronomical Observatory of Japan, Osawa 2-21-1, Mitaka, Tokyo 181-8588, Japan}
\affil{SOKENDAI (The Graduate University for Advanced Studies), 2-21-1 Osawa, Mitaka, Tokyo 181-8588, Japan}
\affil{Department of Astronomy, School of Science, University of Tokyo, Bunkyo, Tokyo 113-0033, Japan}

\author{Kazuhiro~D. Kanagawa}
\affil{Research Center for the Early Universe, Graduate School of Science, University of Tokyo, Bunkyo, Tokyo 113-0033, Japan}

\author{Satoshi Okuzumi}
\affil{Department of Earth and Planetary Sciences, Tokyo Institute of Technology, 2-12-1 Ookayama, Meguro, Tokyo, 152-8551, Japan}

\author{Shigeru Ida}
\affil{Earth-Life Science Institute, Tokyo Institute of Technology, 2-12-1 Ookayama, Meguro, Tokyo 152-8550, Japan}

\author{Catherine Walsh}
\affil{School of Physics and Astronomy, University of Leeds, Leeds, LS2 9JT, UK}

\author{Tom~J. Millar}
\affil{Astrophysics Research Centre, School of Mathematics and Physics, Queen's University Belfast, University Road, Belfast BT7 1NN, UK}

\author{Sanemichi~Z. Takahashi}
\affil{Division of Radio Astronomy, National Astronomical Observatory of Japan, Osawa 2-21-1, Mitaka, Tokyo 181-8588, Japan}
\affil{Division of Liberal Arts, Kogakuin University, 1-24-2 Nishi-Shinjuku, Shinjuku-ku, Tokyo, 163-8677, Japan}

\author{Jun Hashimoto}
\affil{Astrobiology Center, National Institutes of Natural Sciences, 2-21-1 Osawa, Mitaka, Tokyo 181-8588, Japan}

\author{Taichi Uyama}
\affil{Department of Astronomy, School of Science, University of Tokyo, Bunkyo, Tokyo 113-0033, Japan}

\author{Motohide Tamura}
\affil{Department of Astronomy, School of Science, University of Tokyo, Bunkyo, Tokyo 113-0033, Japan}
\affil{Astrobiology Center, National Institutes of Natural Sciences, 2-21-1 Osawa, Mitaka, Tokyo 181-8588, Japan}




\begin{abstract}
We report the detection of an excess in dust continuum emission at 233~GHz (1.3~mm in wavelength) in the protoplanetary disk around TW~Hya revealed through high-sensitivity observations at $\sim$3~au resolution with the Atacama Large Millimeter/submillimeter Array (ALMA).
The sensitivity of the 233~GHz image has been improved by a factor of 3 with regard to that of our previous cycle 3 observations.
The overall structure is mostly axisymmetric, and there are apparent gaps at 25 and 41 au as previously reported. 
The most remarkable new finding is a few au-scale excess emission in the south-west part of the protoplanetary disk.
The excess emission is located at 52 au from the disk center and is 1.5 times brighter than the surrounding protoplanetary disk at a significance of 12$\sigma$.
We performed a visibility fitting to the extracted emission after subtracting the axisymmetric protoplanetary disk emission and found that the inferred size and the total flux density of the excess emission are 4.4$\times$1.0~au and 250~$\mu$Jy, respectively.
The dust mass of the excess emission corresponds to 0.03~$M_\oplus$ if a dust temperature of 18~K is assumed.
Since the excess emission can also be marginally identified in the Band 7 image at almost the same position, the feature is unlikely to be a background source.
The excess emission can be explained by a dust clump accumulated in a small elongated vortex or a massive circumplanetary disk around a Neptune mass forming-planet.
\end{abstract}

\keywords{protoplanetary disks --- stars: individual(TW~Hya)}


\section{Introduction} \label{sec:intro}
It is widely accepted that protoplanetary disks (PPDs) are the birthplace of planets.
Obtaining observational evidence of a forming planet in PPDs is crucial for the understanding of the formation of and the diversity of (exo-)planets \citep{bib:ida2004}.
The detection of a substructure related to a forming planet is a promising way to investigate the planet formation process.
Recent high-resolution observations using ALMA have revealed complex disk substructures, such as multiple, axisymmetric gaps, spiral arms, and lopsided emissions \citep[e.g., DSHARP;][]{bib:andrews2018}, which are likely related to the planet formation process.\par

On the other hand, it is theoretically predicted that the forming planet accretes material from a circumplanetary disk (CPD).
Such CPDs hold the promise of direct detection with sensitive detectors as a localized small-scale substructure in the PPD, and some theoretical models suggest that CPDs should be detectable at millimeter wavelengths \citep{bib:zhu2016}.
Large-scale non-axisymmetric substructure in PPDs has been found at millimeter wavelengths for Oph~IRS~48 \citep[$\sim$60~au in azimuthal extent;][]{bib:vandermarel2013}, HD~142527 \citep[$\sim$150~au;][]{bib:fukagawa2013,bib:casassus2013}, MWC~758 \citep[$\sim$30~au;][]{bib:dong2018}, HD143006 \citep[$\sim$86~au;][]{bib:perez2018}, HD~163296 \citep[$\sim$17~au;][]{bib:isella2018}, and HD~135344B \citep[$\sim$210~au;][]{bib:cazzoletti2018}.
Those lopsided substructures are interpreted as the accumulation of dust particles due to a large-scale gas vortex in the PPD.
On the other hand, au-scale substructures or non-axisymmetric components, which are expected to be the signatures of CPDs, have not yet been discovered at these wavelengths \citep{bib:isella2014}.\par

TW~Hya is the nearest T Tauri star with a distance of 59.5~pc \citep{bib:gaia2016}.
The stellar mass is 0.8~$M_\sun$ and the stellar age is 10~Myr \citep{bib:andrews2012}.
The disk orientation is almost face-on with an inclination of 7$\degr$ \citep{bib:qi2004}.
High-resolution observations with ALMA have resolved multiple axisymmetric gaps; in particular, deep gaps at $\sim$25 and 41~au \citep{bib:andrews2016,bib:tsukagoshi2016,bib:huang2018}.
Evidence for non-axisymmetry or small-scale substructure has not yet been reported for the PPD around TW~Hya.\par

In this paper, we report the results of our high-sensitivity observations at Band~6 using ALMA and the first finding of an au-scale substructure at a radius of 52~au in the PPD around TW~Hya.
We describe our observations and data reduction in \S \ref{sec:obs}.
In \S \ref{sec:results}, we show the resulting images of the Band~6 continuum emission, and the finding of the au-scale substructure in the TW~Hya disk is presented.
In \S \ref{sec:discussion}, we discuss the detailed structures and the expected origin of the substructure.

\section{Observations} \label{sec:obs}
\subsection{Sensitive ALMA Observation at Band 6}
Our 233 GHz (1.3~mm) continuum observations at Band~6 toward TW~Hya were carried out using ALMA on 15 May 2017 with array configuration C40-5 (2016.1.00842.S) and in the period from 20 to 25 November 2017 with C43-8 (2017.1.00520.S).
The total on-source integration times were $\sim$12 and $\sim200$~minutes, respectively.
The correlator was configured to detect dual polarizations in four spectral windows with a bandwidth of 1.875~GHz each, resulting a bandwidth of 7.5~GHz in total.
The four spectral windows were tuned to detect continuum emission centered at 225, 227, 239, and 241~GHz.
The phase fluctuations of the complex gain due to atmospheric noise were calibrated by observing quasars J1051-3138 or J1103-3251.
Quasars J1107-4449, J1058+0133, or J1037-2934 were used for the calibration of the bandpass characteristics and
flux scales were determined by observing J1058+0133, J1107-4449, or J1037-2934.\par

The observed visibilities were reduced and calibrated using the Common Astronomical Software Application (CASA) package \citep{bib:mcmullin2007}.
The initial flagging of the visibilities and the calibrations for the bandpass characteristics, complex gain, and flux scaling were performed using the pipeline scripts provided by ALMA.
After flagging the bad data, the corrected data were concatenated and imaged by CLEAN.
The CLEAN map was created by adopting Briggs weighting with a robust parameter of 0.5.
We also employed the multiscale CLEAN with scale parameters of [0, 42, 126]~mas for better reconstruction of extended emission.
After that, self-calibration was applied for the concatenated data set.
The spatial resolution of the CLEANed image was 46.88$\times$41.56 mas, full width at half maximum (FWHM) with a position angle of $-78.9^\circ$, corresponding to 2.79$\times$2.48~au.
The 1$\sigma$ root-mean-square (rms) noise level achieved was 9.1 $\mu$Jy beam$^{-1}$.

\subsection{Data Reduction of the Band 7 Archive Data}
To validate the new results revealed by our Band~6 observation, we have also analysed eight sets of ALMA archival data at Band~7 from Cycle~3.
The highest resolution data obtained by \citet{bib:andrews2016} is included, which is taken at the beginning of December 2015, two years before our observations.
Data reduction has been done in the same manner as for our Band~6 observations described above.
Line free channels were used for making the map of continuum emission at Band~7.
The center frequency of the concatenated data was 325.4~GHz.
The calibrated data were concatenated and imaged by CLEAN with an iterative self-calibration.
For the imaging, we adopt Briggs weighting with a robust parameter of -1.0 so that the synthesized beam can be described well by a single Guassian function.
Multiscale CLEAN with scale parameters of [0, 30, 90]~mas was employed.
The final CLEANed image had a synthesized beam size of 36.4$\times$28.9~mas (2.2$\times$1.7~au) with a position angle of 69.9$^\circ$.
To reduce phase noise, we have smoothed the final CLEANed image to 50 mas resolution using the {\it imsmooth} task on CASA.
The rms noise level of the resultant image is 27.7~$\mu$Jy beam$^{-1}$.\par

\section{Results} \label{sec:results}
\begin{figure*}
\begin{center}
    \includegraphics[width=\textwidth]{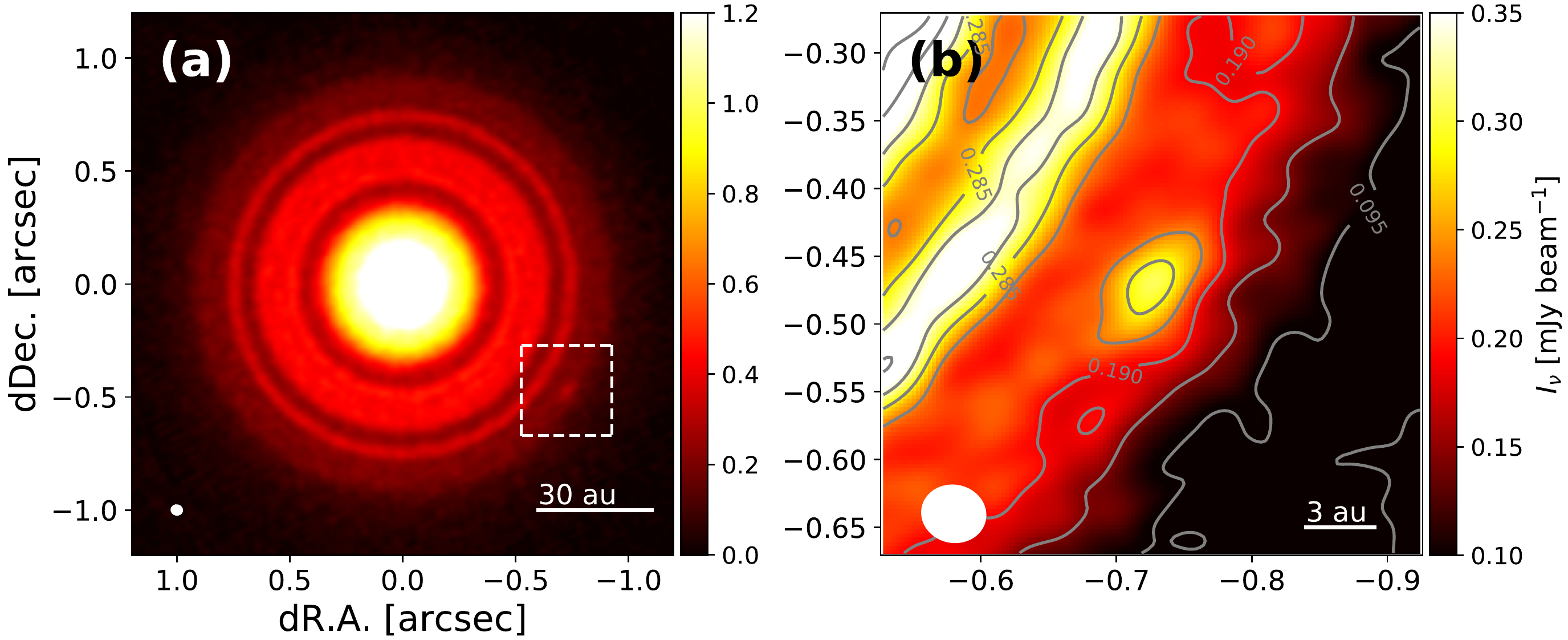}
    \caption{233 GHz continuum maps. The white ellipse at the bottom left corner of each panel indicates the beamsize of the synthesised images. (a) Overall distribution of the 233~GHz continuum emission. (b) Close-up view of the $0\farcs4\times0\farcs4$ box including the emission feature (white box in the main panel). The contour interval is 5$\sigma$, where 1$\sigma=$9.1 $\mu$Jy beam$^{-1}$.}
    \label{fig:fig1}
\end{center}
\end{figure*}

Figure \ref{fig:fig1}a shows the global distribution of the 233 GHz continuum emission.
The total flux density from the PPD and the known axisymmetric features, gaps at 25 and 41~au, are consistent with previous works \citep{bib:andrews2016,bib:tsukagoshi2016,bib:huang2018}.\par

\begin{figure*}
\begin{center}
   \includegraphics[width=\textwidth]{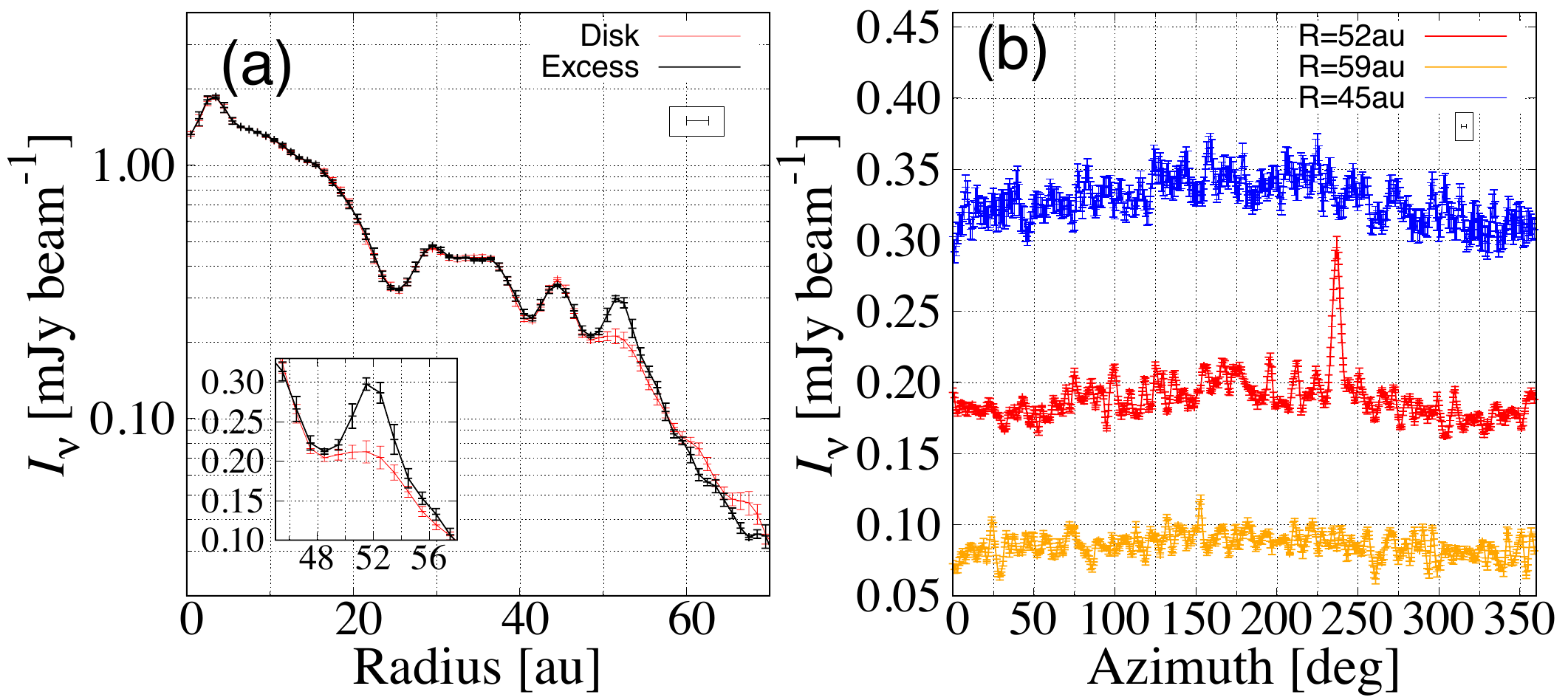}
    \caption{Deprojected radial and azimuthal profiles of the 233~GHz continuum emission. The bar at the top-right corner in the box shows the FWHM of the synthesized beam. The error bars are determined from the standard error through the averaging. (a) Radial profile running through the emission feature (P.A.$=236^\circ$--$238^\circ$) is shown in black. The red line represents the radial profile averaged over the neighbourhood of the emission feature (P.A.=222--236$^\circ$ and 238--252$^\circ$). The panel at the bottom left corner of the left panel shows the close-up view of the radial profile around 52~au on a linear scale. (b) Azimuthal profiles of the emission at radii of 52, 59, and 45~au are shown in red, yellow and blue, respectively.}\label{fig:fig2}
\end{center}
\end{figure*}

In the south-west part of the PPD, the resolved excess of emission is discovered at a radius of 52~au, as shown in Fig. \ref{fig:fig1}b.
The peak intensity at this feature is measured to be 308.4 $\mu$Jy beam$^{-1}$, corresponding to signal-to-noise (SN) ratio of 34.
The excess of emission is significant in comparison to the surrounding background emission of the PPD.
The excess is clearly detected both in the radial and in the azimuthal profiles as shown in Figure \ref{fig:fig2}.
The average of the surrounding emission is measured to be 197.1 $\mu$Jy beam$^{-1}$ from position angles of 226--230$^\circ$ and 244--248$^\circ$; hence the excess is calculated to be 111.5$\pm$9.1~$\mu$Jy beam$^{-1}$ ($\sim12\sigma$), i.e., the emission feature is 1.5 times brighter than the surrounding PPD.
There is no clear excess of emission above 3$\sigma$ in this radial region of the PPD except for the emission feature.
We also note that no clear gap-like structure is found at 52~au from the central star.\par


Since the emission from the PPD is almost wholly axisymmetric, the emission feature is extracted by subtracting the axisymmetric background emission in the visibility (i.e., $u$-$v$) domain.
The size, position and total flux density of the emission feature are measured by model fitting in the $u$-$v$ domain using the residual visibilities.\par

\begin{figure*}
\begin{center}
    \includegraphics[width=\textwidth]{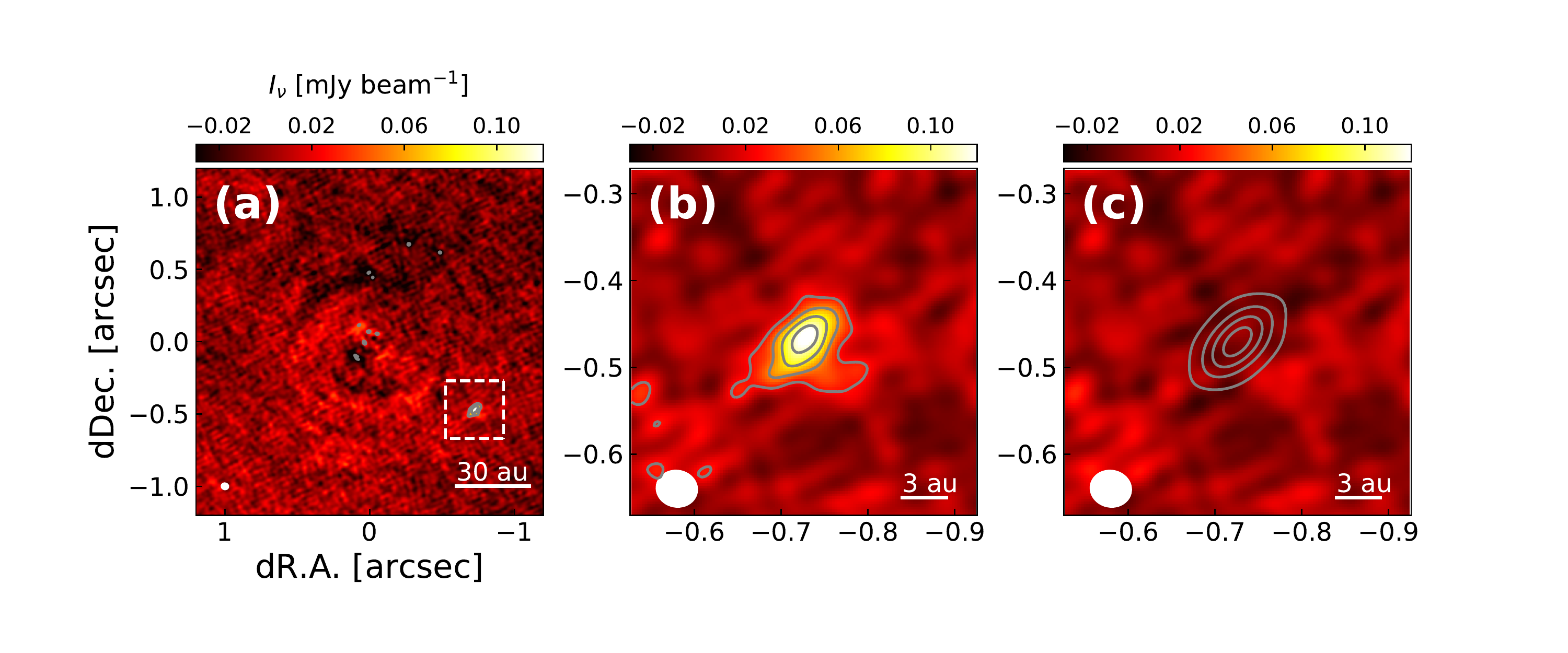}
    \caption{233~GHz images after the extraction of the emission feature. (a) CLEANed image of the residual emission reconstructed from the visibilities obtained by subtracting the axisymmetric emission in the $u$-$v$ plane. (b) Close-up view of the $0.4^"\times0.4^"$ box of the residual image (white box in the main panel a). The grey contour starts at $\pm$3$\sigma$ with an interval of 3$\sigma$. (c) \replaced{Result of the 2D Gaussian fitting to the emission feature is shown in colour. The grey contour is the same as that in panel b.}{Result of the 2D Gaussian fitting to the emission feature (contour) and the difference between the fitted Gaussian and the extracted emission feature (color). The grey contour starts at $\pm3\sigma$ with an interval of 3$\sigma$.}}\label{fig:fitblob}
\end{center}
\end{figure*}

The subtraction process works well as shown in Figure \ref{fig:fitblob}a and b.
To deduce the structure of the emission feature, we performed a least squares fitting to the subtracted visibilities using the {\it uvmodelfit} task installed on CASA.
A single component described by a Gaussian function was assumed for the model.
The modeled visibilities determined by the fitting were converted to an image using CLEAN with the same parameters as for the original image, and the resulting image is shown in Fig. \ref{fig:fitblob}c.
From the fitting, the peak position of the emission feature is measured to be ($\Delta\alpha$, $\Delta\delta$)$=$($-726\pm1$, $-471\pm1$)~mas from the central star, corresponding to ($-43.2\pm0.1$, $-28.0\pm0.1$)~au or a radius of 51.5$\pm$0.1~au.
The FWHMs of major and minor axes of the fitted Gaussian are measured to be (74.7$\pm$3.3)$\times$(16.2$\pm$3.2) mas with a position angle of $-38.3^\circ\pm2.1^\circ$, corresponding to (4.4$\pm$0.2)$\times$(1.0$\pm$0.2)~au.
The total intensity of the fitted Gaussian is 250$\pm$5 $\mu$Jy.\par

The brightness temperature of the emission feature can be converted from the total flux density.
Assuming that the millimeter emission of the PPD behind the resolved feature is uniform, the total flux density of the background inside the feature area is estimated to be 122~$\mu$Jy from the average of the surrounding emission, and thus the summation of the total flux density inside this area is 372~$\mu$Jy. 
Therefore, the brightness temperature at the position of the emission feature is converted using the Planck function to be 11.6~K.
The brightness temperature is \added{nearly equal to or} less than the temperature of the dust disk at 52~au, where $\sim$14--18~K is expected from the midplane temperature profile \citep{bib:andrews2016,bib:tsukagoshi2016,bib:huang2018}, indicating that the emission feature \replaced{at the radius of 52~au is}{may be partially} optically thin.\par

\added{Assuming optically thin conditions,} the mass of the emission feature is estimated from the equation 
\begin{equation}\label{eq:dustmass}
M_\mathrm{dust} = \frac{F_\nu d^2}{\kappa_\nu B(T_\mathrm{dust})},
\end{equation}
where $F_\nu$ is the integrated flux density, $d$ is the distance to the source, and $B(T_\mathrm{dust})$ is the Planck function for the dust temperature $T_\mathrm{dust}$.
We employ a dust mass opacity coefficient $\kappa_\nu$ of 2.3 cm$^2$ g$^{-1}$ at a frequency of 233~GHz, which is determined from $\kappa_\nu$ of 2.8 cm$^2$ g$^{-1}$ at a wavelength of 870~$\mu$m \citep{bib:andrews2012} and $\beta$, the power-law index of $\kappa_\nu$, of $\sim0.5$.
Assuming $T_\mathrm{dust}=18$~K \citep{bib:huang2018}, the total dust mass of the emission feature is (2.83$\pm$0.06)$\times10^{-2}$ $M_\oplus$, where $M_\oplus$ indicates the Earth mass.
\added{It should be noted that the estimated mass depends on the uncertainty of the opacity and could be a lower limit because the emission may not be completely optically thin.}\par

\begin{figure*}
\begin{center}
	\includegraphics[width=\textwidth]{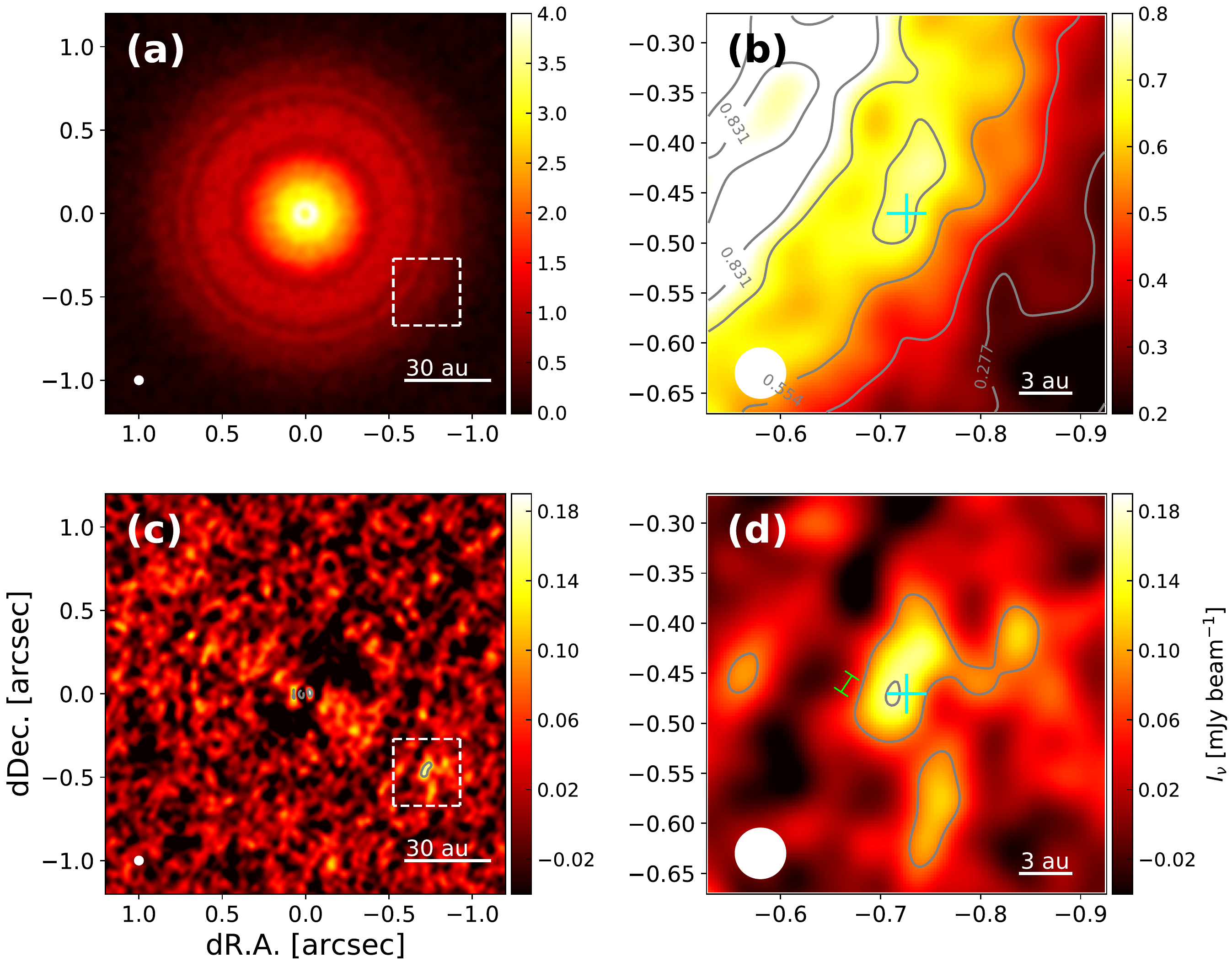}
    \caption{325 GHz continuum map at 50~mas resolution made from the publicly-available data. The white ellipse at the bottom left corner in each panel indicates the beamsize of the images. (a) Overall distribution of the emission. (b) Close up view of the box with white dotted lines in the panel (a). The contour interval is 5$\sigma$, where 1$\sigma=$27.7 $\mu$Jy beam$^{-1}$. The cross in cyan indicates the peak position of the emission feature identified in our 233~GHz image. (c) Residual emission after the subtraction of the axisymmetric component from the overall image. The $\pm$5$\sigma$ contour is shown. (d) Close-up view of the box with white dotted lines in the panel (c). The contour starts at $\pm$3$\sigma$ with an interval of 3$\sigma$. The green line on the left of the emission feature indicates the expected Keplerian motion at 52~au for 2~yrs.\label{fig:b7image}}
\end{center}
\end{figure*}

The emission feature can also be seen in a high-resolution 325~GHz ALMA image within close proximity to the location of our 233~GHz emission feature.
The 325~GHz continuum image we reconstructed from the archive data is shown in Figure \ref{fig:b7image}a and b, and the subtracted image made by using the same procedure as done for our 233~GHz data is shown in Figure \ref{fig:b7image}c and d.
It is clear that there is a local emission peak near the position of the emission feature that we found in the 233~GHz map, while the emission seems to be azimuthally elongated.
The positional offset between the emission feature at Band~6 and the residual emission at Band~7 is much less than the proper motion of the TW~Hya system.
If the excess emission is a background source, the positional offset must be 136~mas for 2~yrs in the R.A. direction according to the proper motion of the TW~Hya system \citep[($-68.225$, $-13.934$)~mas~yr$^{-1}$;][]{bib:gaia2016}.
The positional deviation of the residual emission at Band~7 is within only $\sim$50~mas with respect to the emission feature at Band~6, as shown in Figure \ref{fig:b7image}.
Therefore, we conclude that the emission feature we found is situated in the PPD and is likely orbiting the central star.
\replaced{It is unclear whether or not there is a positional offset due to the Keplerian motion, because it is expected to be only $\sim$26~mas for 2~years at 52~au (see Fig. \ref{fig:b7image}d).}{Due to the significant phase noise of the Band 7 data, we conservatively conclude that it is unclear whether there is a positional offset due to the Keplerian motion for 2~years. (see Fig. \ref{fig:b7image}d).}

Lastly, we also note that the millimeter flux density fluctuates azimuthally (see the profile at 45~au in Fig.~\ref{fig:fig2}b).
The fluctuation might be related to a moving surface brightness asymmetry which is probably due to disk shadow \citep{bib:debes2017}, but we do not focus on this structure in this paper.

\section{Discussion}\label{sec:discussion}
We have found a few au-scale, elongated emission feature in the PPD of TW~Hya.
Similar asymmetric structures have been found in several other PPDs as listed in \S\ref{sec:intro}, but the feature in the TW~Hya disk is the smallest ever discovered.

In many PPDs, the asymmetric features are interpreted as the dust particles trapped in a gas vortex \citep[e.g., ][]{bib:raettig2015}.
The morphology of the emission feature we found may also be interpreted along this line.
The width of the gas vortex is limited to several times the gas scale height.
The radial half width of the emission feature is only $\sim$0.5~au and is much smaller than the expected gas scale height at this radius which is $\sim$4--5~au \citep{bib:andrews2012}.
The emission feature is azimuthally elongated and the ratio of the azimuthal to the radial widths, or aspect ratio, is $\sim$4.
The shape of gas vortices has been investigated by several authors \citep{bib:lesur2009,bib:richard2013}, and the aspect ratio of stable vortices is of the order of $\sim$several.
The dust particles trapped in a gas vortex is more concentrated at the vortex center, but the aspect ratio of the dust distribution is similar to that of gas \citep{bib:lyra2013}.

The mean surface density of the emission feature is expected to be $\sim$2 times higher than that of the surrounding PPD under the assumption that the temperature at 52~au is constant at 18~K \citep{bib:huang2018}.
If the overall dust-to-gas mass ratio of the TW~Hya disk is $\sim$100, the excess emission region may have a dust-to-gas mass ratio of around 50.
Such a small overdensity may be realized even with a weak gas vortex.

The gas disk may be full of such ``weak vortices'' if the PPD is moderately turbulent.
A chain of vortices in the same radial location will merge into one vortex \citep[e.g., ][]{bib:ono2018}, but there may be more vortices at different radii.
In this sense, the emission feature could be a ``tip of the iceberg'' of even smaller dust concentrations.

An alternative scenario that may explain why there is only {\it one} emission feature is the existence of a planet.
If there is a planet accreting gas and dust from the surrounding PPD \citep[e.g., ][]{bib:pollack1996,bib:canup2002}, the temperature and the density may increase locally \citep[e.g., ][]{bib:owen2014}
\added{
and a CPD may be formed around the planet.
The emission feature may be the remnant of the accretion onto the planet and/or the CPD.
}

\added{
We first make a rough estimate of the planet mass using the radial half width of the emission feature ($\sim 0.5$~au), which may be interpreted as the maximum radius of the putative CPD.  The size of the CPD is considered to be several times smaller than the Hill radius $r_{\mathrm{H}}= r_\mathrm{p}(M_\mathrm{pl}/3M_{\ast})^{1/3}$, where, $r_\mathrm{p}$ is the orbital radius of the planet, $M_\mathrm{p}$ is the planet mass, and $M_{\ast}$ is the mass of the central star.  
The exact size of the CPD has been a topic of active debate recently.  It has been considered as $\sim r_{\mathrm{H}}/3$ \citep{bib:quillen1998,bib:ayliffe2009} while recent simulations indicate that it may be $\sim r_{\mathrm{H}}/10$ or smaller for a low-mass planet like Neptune \citep[e.g., ][]{bib:wang2014,bib:ormel2015,bib:szulagyi2018b}.
The size of the emission feature corresponds to $r_{\mathrm{H}}/3$ for a Neptune mass planet and $r_{\mathrm{H}}/10$ for a 30 Neptune mass planet.

Two observational evidences prefer a lower mass planet.  
Firstly, the existence of a planet more massive than $\sim$1--2 Jovian mass ($=$20--40 Neptune mass) at $\sim 50$~au from the central star is ruled out by 3.8~$\mu$m (L$^{\prime}$-band) observations \citep{bib:ruane2017}.  
Secondly, we do not observe any gap structure at $r\sim 50$~au.
This requires a low planet mass and/or high viscosity \citep[e.g., ][]{bib:kanagawa2017}.
\citet{bib:dipierro2017} discusses that a planet with a mass of $\lesssim \mathrm{several} \times 10^{-5}~M_{\mathrm{\ast}}$, roughly corresponding to Neptune mass, does not form a significant gap in both the gas and dust distribution, while such low mass planets may still open a gap in a low viscosity ($\alpha \lesssim 10^{-4}$) environment \citep[e.g., ][]{bib:dong2017,bib:dong2018b}.  
Since only loose upper limits on the $\alpha$-parameter of viscosity is given observationally \citep[$\alpha \lesssim \mathrm{several} \times 10^{-3}$; ][]{bib:teague2016,bib:flaherty2018}, we consider $\sim$ 1~Neptune mass as an upper limit for the planet mass.
}
\deleted{
If the radial half width of the emission feature is 1/3 of the Hill radius $r_\mathrm{H}$ of a planet, the mass of the unseen planet surrounded by the CPD is estimated to be Neptune mass.
Here, $r_\mathrm{H} = r_\mathrm{p}(M_\mathrm{pl}/3M_{\ast})^{1/3}$, where $r_\mathrm{p}$ is the orbit of the planet, $M_\mathrm{p}$ is the planet mass, and $M_{\ast}$ is the mass of the central star.
Since the ratio of the estimated planet mass to the stellar mass reaches of order of 10$^{-5}$, it is consistent with the fact that we do not find any gap structure at the radius of the emission feature \citep{bib:dipierro2017}.
}

\begin{figure}
\begin{center}
	\includegraphics[width=0.5\textwidth]{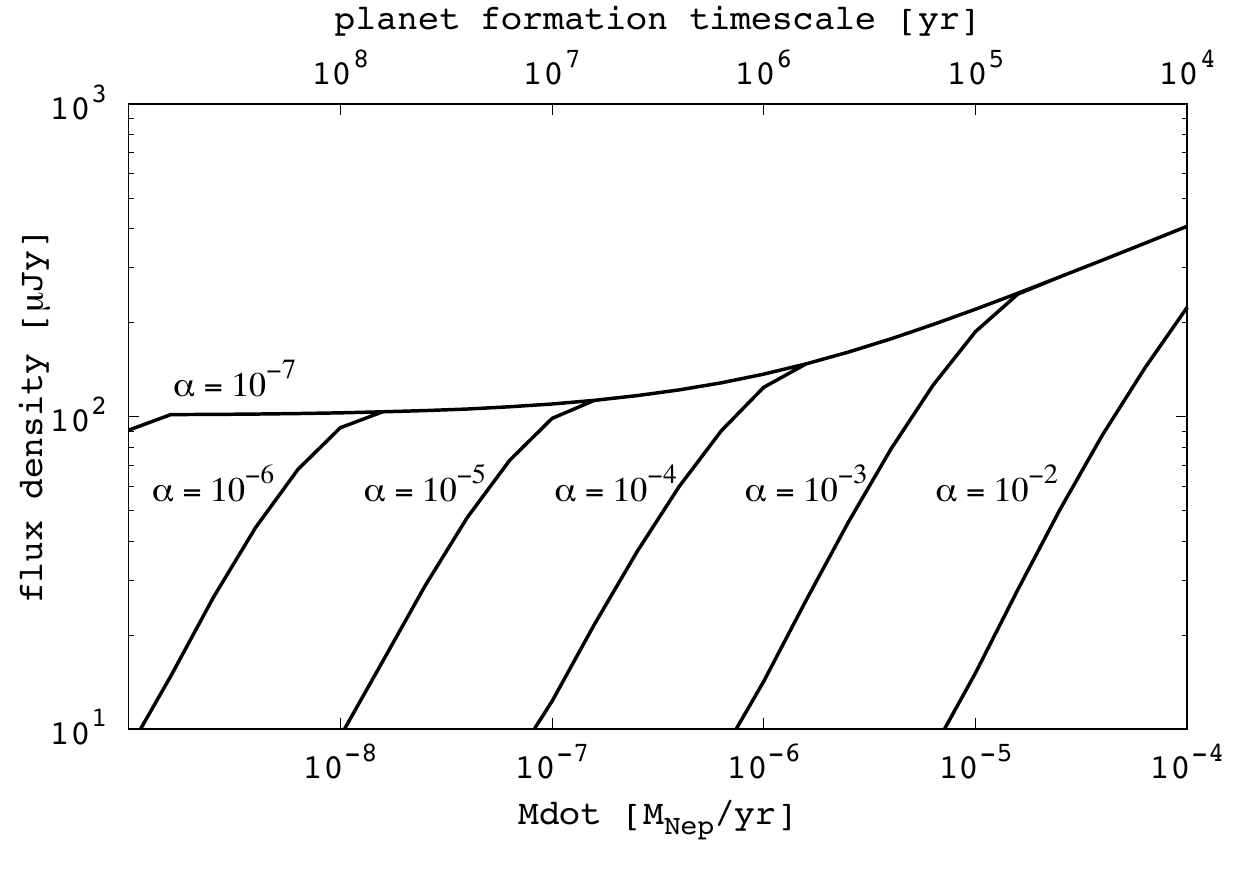}
	    \caption{Flux density of the model CPDs for a range of viscous parameters $\alpha$ and mass accretion rates $\dot{M}$. The horizontal axis shows the mass accretion rate in the unit of Neptune masses per year. The upper horizontal axis indicates the planet formation timescale which is calculated by $M_{pl}/\dot{M}$ assuming a constant mass accretion rate. Different solid lines are for calculations with different viscous parameters.\label{fig:CPD_alpha_Mdot}}
\end{center}
\end{figure}

However, it should be noted that the observed emission feature may not be fully accounted for by emission from a CPD.
We use a simple model by \citet{bib:zhu2016} to estimate the flux density at millimeter wavelengths from the CPD 
\added{with a radius of $r_{\mathrm{H}}/3$}
around a Neptune-mass planet.
\deleted{Here, we assume that the CPD is circular with a radius of $0.5$~au around the planet, which is $1/3$ of the Hill radius.}
Figure \ref{fig:CPD_alpha_Mdot} shows the flux density of the modeled CPDs for a range of viscous parameters $\alpha$ and mass accretion rates onto the planet $\dot{M}$.
If the mass accretion rate is $10^{-7}~M_{\mathrm{Nep}}$ per year, the flux density of the CPD is no larger than $\sim 100~\mu$Jy.
\added{We consider this as an upper limit since the CPD around a low-mass planet may be much smaller in size, resulting in a much smaller emitting area.}
This value may be compared with the observed emission within a circle with a diameter of 1~au, 
\added{which is $\sim 60~\mu$Jy as the total emission is $\sim 250~\mu$Jy within $ 4.4 \times 1.0$~au.
} 
\deleted{The contribution to the flux density from the inner $1 \times 1$~au region of the emission feature is $\sim 60~\mu$Jy since the emission has a flux density of $\sim 250~\mu$Jy within $ 4.4 \times 1.0$~au.}
Therefore, a part of the excess emission may come from the CPD while the entire emission may be accounted for by the overdensity due to a surrounding envelope-like structure of accreting material around the system of the planet and the CPD.
\deleted{around a hypothetical Neptune mass planet.
The entire 250~$\mu$Jy flux density of the emission feature may be explained by a overdensity due to a surrounding envelope-like structure of accreting material around the system of the planet and the CPD.
}
\deleted{
It should also be noted that the flow structure around a low mass planet, and even the existence of the CPD, is still under debate
}

\added{In short, in the case of the planet scenario, the allowed planet mass range may be limited to 1 Neptune mass or less.  The CPD around the putative planet may account for a part of the excess emission, while it is not possible to explain the entire feature.
We note that \citet{bib:dong2018b} suggested a $\sim$2 Neptune mass planet at 45~au as a cause of gap structures.  The putative planet at the location of the emission feature may not coexist with this planet since the orbits are too close together to form a stable system.}

With observations at millimeter wavelengths only, it seems difficult to judge whether the hypothetical planet actually exists at the location of the emission feature.
The simple CPD model described above suggests that the millimeter emission from the CPD is similar to that of dust concentration within a gas vortex, making it difficult to determine whether the planet exists or not.
The most prominent feature of an accreting planet is 
\replaced{that the materials in the close vicinity of the planet may be heated up to $\gtrsim 1000$~K due to the release of the gravitational energy. 
}{that the material heated up to $\gtrsim 1000$~K in the close vicinity of the planet.}
Therefore, direct imaging observations at infrared wavelengths or detection of an accretion signature in emission lines may be critical to prove (or rule out) the existence of a planet.
\added{Further, high angular resolution observations of molecular line emission may reveal the kinematics of the emission feature, thus providing further evidence for the presence or otherwise of a vortex.}

\deleted{
The existence of a planet more massive than $\sim$1--2 Jovian mass at $\sim 50$~au from the central star is ruled out by 3.8~$\mu$m (L$^{\prime}$-band) observations \citep{bib:ruane2017}, while a Neptune mass planet would have remained undetected.
}

\acknowledgments
This paper makes use of the following ALMA data: ADS/JAO.ALMA\#2017.1.00520.S and 2016.1.00842.S.
We also use the following public ALMA archive data: ADS/JAO.ALMA\#2015.1.00308.S, 2015.1.00686.S, 2016.1.00229.S, 2016.1.00311.S, 2016.1.00440.S, 2016.1.00464.S, 2016.1.00629.S and 2016.1.01495.S.
ALMA is a partnership of ESO (representing its member states), NSF (USA) and NINS (Japan), together with NRC (Canada), NSC and ASIAA (Taiwan), and KASI (Republic of Korea), in cooperation with the Republic of Chile.
The Joint ALMA Observatory is operated by ESO, AUI/NRAO and NAOJ. A part of the data analysis was carried out on the common-use data analysis computer system at the Astronomy Data Center of NAOJ.
This work is partially supported by JSPS KAKENHI grant number 17K14244.
TJM thanks STFC for support under grant reference ST/P000321/1. CW acknowledges financial support from STFC (grant reference ST/R000549/1) and the University of Leeds.

%

\vspace{5mm}
\facilities{ALMA}


\software{astropy, 
          CASA, 
          }

\end{document}